\documentclass[aps,prb,showpacs,reprint,superscriptaddress,floatfix]{revtex4-1}

\usepackage{graphicx}
\usepackage{color}
\usepackage{amsmath}
\usepackage{bbm}
\usepackage{amssymb}

\usepackage{dsfont}

\usepackage[utf8]{inputenc}	% kodowanie UTF-8
\usepackage[T1]{fontenc}

\input epsf

\newcommand{\Lande}{Land\'e}           % Landé
\newcommand{\RLZ}{RLZ}                 % Roth-Lax-Zwerdling
\newcommand{\gstar}{g^\ast}
\newcommand{\tgstar}{\mathbf{g}^\ast} 
\newcommand{\valpha}{\boldsymbol{\alpha}_R}
\renewcommand{\H}{\mathbf{H}}
\begin{document}

%Authors
\title{Enhancement and anisotropy of electron \Lande\ factor due to spin-orbit interaction in semiconductor nanowires}

\author{Julian Czarnecki}
\email{jczarnecki@student.agh.edu.pl}
\affiliation{AGH University of Krakow, Faculty of Physics and Applied Computer Science, al. A. Mickiewicza 30, 30-059 Krakow, Poland}

\author{Andrea Bertoni}
\email{andrea.bertoni@nano.cnr.it}
\affiliation{CNR-NANO S3, Istituto Nanoscienze, Via Campi 213/a, 41125 Modena, Italy}

\author{Guido Goldoni}
\email{guido.goldoni@unimore.it}
\affiliation{Department of Physics, Informatics and Mathematics, University of Modena and Reggio Emilia, via Campi 213/a, 41125 Modena, Italy}
\affiliation{CNR-NANO S3, Istituto Nanoscienze, Via Campi 213/a, 41125 Modena, Italy}

\author{Paweł W\'ojcik}
\email{pawel.wojcik@fis.agh.edu.pl}
\affiliation{AGH University of Krakow, Faculty of Physics and Applied Computer Science, al. A. Mickiewicza 30, 30-059 Krakow, Poland}

\date{February 2023}

\begin{abstract}
We investigate the effective \Lande\ factor in semiconductor nanowires with strong Rashba spin-orbit coupling. Using the $\mathbf{k}\cdot\mathbf{p}$ theory and the envelope function approach we derive a conduction band Hamiltonian where $\gstar$ is explicitly related to the spin-orbit coupling contants $\valpha$. Our model includes orbital effects from the Rashba spin-orbit term, leading to a significant enhancement of the effective \Lande\ factor which is naturally anisotropic. For nanowires based on the low-gap, high spin-orbit coupled material InSb, we investigate the anisotropy of the effective \Lande\ factor with respect to the magnetic field direction, exposing a twofold symmetry for the bottom gate architecture. The anisotropy results from the competition between the localization of the envelope function and the spin polarization of the electronic state, both determined by the magnetic field direction.

\end{abstract}

\maketitle

\section{Introduction}

Semiconductor nanowires (NWs) continue to attract significant interest due to the abundance of physical phenomena observed in such nanostructures, as well as the wealth of potential applications, including optoelectronics,\cite{Reimer,Stettner,Li,Czaban2009} quantum computing,\cite{NadjPerge,Frolov,Schroer} or spintronics.\cite{Pribiag,Miladi,NadjPergePRL,Wojcik2014} Applications in spintronics are largely driven by the spin-orbit (SO) interaction, which -- in low energy gap semiconductors, such as InAs or InSb -- is sufficiently strong to enable electrical control of the electron spin. In general, the SO interaction originates from the lack of the inversion symmetry, which could be an intrinsic feature of the crystallographic structure (Dresselhaus SO coupling~\cite{Dresselhaus}) or induced by the asymmetry of the confinement potential (Rashba SO coupling\cite{Rashba}). The latter has the essential advantage of being tunable by external fields, e.g., using gates attached to the nanostructures, as predicted theoretically\cite{Campos,Kokurin2015,Kokurin2014,Wojcik2018,Wojcik2021,Escribano,Wojcik2019,Furthmeier} and demonstrated in recent experiments.\cite{vanWeperen2015,Kammhuber2017,Dhara2009,Scherubl2016,Liang2012}  

The significant progress in heteroepitaxy, which has been made over the last decade, enables the growth of a thin superconducting layer on the surface of semiconductor.\cite{gazibegovic_epitaxy_2017,krogstrup_epitaxy_2015,chang_hard_2015,kjaergaard_quantized_2016} In this respect hybrid NWs with a large SO interaction are recently intensively studied as the basic building blocks for topological quantum computing based on Majorana zero modes.\cite{mourik_signatures_2012,deng_anomalous_2012,albrecht_exponential_2016,zhang_quantized_2017,finck_anomalous_2013} These exotic states are formed at the ends of NWs when the system becomes spinless, which is achieved in experiments by applying a magnetic field and the corresponding spin Zeeman effect.\cite{mourik_signatures_2012} The induced topological gap strongly depends on the strength of the SO coupling and the energy of the Zeeman splitting,\cite{oreg_helical_2010,sau_generic_2010,Lutchyn} usually expressed in terms of a linear response to the magnetic field with a proportionality constant $\gstar$  -- the effective \Lande\ factor. In other words, $\gstar$ determines the strength of the magnetic field required to trigger the system into the topological phase. For this reason, it is desirable to make it as large as possible, as the magnetic field needed for the topological transition is required to be lower than the critical magnetic field of the superconducting shell.\cite{albrecht_exponential_2016}

In semiconducting materials $\gstar$ is significantly different from the free-electron \Lande\ factor $g_0$, due to coupling between the valence and the conduction band. In the second-order perturbation $\mathbf{k}\cdot\mathbf{p}$ theory it leads to the Roth-Lax-Zwerdling (\RLZ) formula,\cite{Roth-Lax} which for low gap semiconductors gives $\gstar \ll g_0$, e.g. $ \gstar \approx -49$ for InSb. In particular, for semiconductor nanostructures the \RLZ\ formula predicts \emph{a reduction} of the effective \Lande\ factor with respect to the bulk value,\cite{Lommer,Kiselev,Gawarecki2020} as the subband confinement increases the energy gap, which is inversely proportional to $\gstar$.\cite{Roth-Lax} However, unexpectedly, recent experiments in NWs based on InAs and InSb exhibit opposite behaviour - the extracted $\gstar$ is up to three times larger than the bulk value.\cite{Schroer,vanWeperen2013,Marcus2018} Furthermore, in Ref.~\onlinecite{Marcus2018} a step like evolution of $\gstar$ has been reported as a function of the gate voltage.  
It has been recently proposed that this surprising behaviour arises from the $\mathbf{L}\cdot \mathbf{S}$ coupling, which for higher subbands (characterized by the large orbital momentum) leads to the enhancement of $\gstar$ by about one order of magnitude.\cite{Winkler2017}

In this paper we develop a full $8 \times 8 ~ \textbf{k} \cdot \textbf{p}$ theory of  the effective \Lande\ factor in semiconductor NWs which takes into account the orbital effects in the SO coupling terms induced by an external magnetic field of arbitrary direction. For a nanowire based on the low-gap, strongly SO coupled material InSb, we performed fully self-consistent calculations taking into account on equal footing orbital and Zeeman effects of the applied magnetic field, SO coupling and the electrostatic environment. We demonstrate that the orbital contribution to $\gstar$ ensuing from the SO interaction may overcome the bulk contribution, leading to the enhancement of the effective \Lande\ factor by an order of magnitude, even for the lowest subband, the one usually considered in Majorana experiments. Finally, we also evaluate the anisotropy of the SO-induced \Lande\ factor with respect to the magnetic field rotated in different planes. Our results qualitatively agree with recent experiments\cite{Schroer,vanWeperen2013,Marcus2018} reproducing the enhancement of $g^*$ and its anisotropy.

The paper is organized as follows. In Sec. II A, the \Lande\ factor is derived from the $8 \times 8 ~ \textbf{k} \cdot \textbf{p}$ model within the envelope function approximation. Details on the numerical method are given in Sec. II B. Sec. III contains results of our calculations for homogeneous InSb NWs and their discussion with respect to recent experiments. Sec. IV summarizes our results.

\section{Theoretical model}

Below we shall derive a $\textbf{k} \cdot \textbf{p}$ formulation of the \Lande\ factor in semicondutor NWs. We shall specifically consider a homogeneous InSb, with hexagonal cross section, grown in the zincblede crystallographic structure along the [111] direction. This particular orientation preserves the crystal inversion symmetry, resulting in the reduction of the Dresselhaus SO coupling term.\cite{Dresselhaus,vanWeperen2015}

\begin{figure}[!htp]
    \centering
    \includegraphics[width = 0.5 \textwidth]{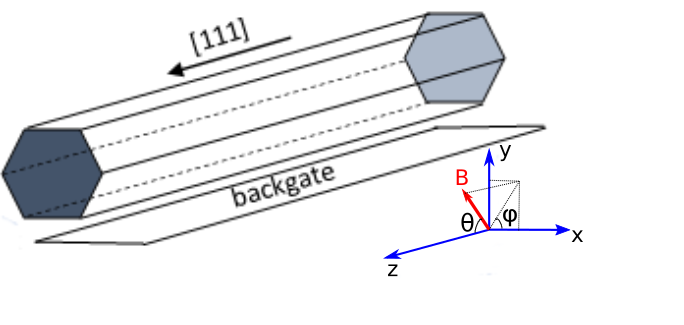}
    \caption{ Schematic illustration of a NW with a bottom gate together with a coordinate system with the magnetic field direction determined by two angles $\theta$ and $\varphi$.}
    \label{fig1}
\end{figure}

The system is subjected to a uniform external magnetic field with intensity $B$.
The direction of the applied magnetic field with respect to the NW axis is determined by the angles $\theta$, between the field and the NW axis ($z$), and $\varphi$, between the $x$ axis (oriented along the corner-corner direction) and the projection of the field on the $xy$ plane -- see Fig.~\ref{fig1}. Hence, 
\begin{equation}
    \begin{split}
        \textbf{B} & = [B_x,B_y,B_z]^T \\
        & =  B \left [\sin(\theta)\cos(\varphi), \sin(\theta)\sin(\varphi),\cos(\theta) \right ]^T\,.
    \end{split}
\end{equation}

\vspace{0.5cm}
We adopt the symmetric vector potential 
\begin{equation}
\label{eq:vec_pot}
\textbf{A}(\textbf{r}) = \left [ -\frac{yB_z}{2}, \frac{xB_z}{2}, yB_x - xB_y \right ]^T ,
\end{equation}
and assume that the back gate is attached directly to the bottom facet of NW, generating an electric field in the $xy$ plane. Although in real experiments a dielectric layer separating the NW from the gate is usually used, it plays a role of screening for the electric field. Hence, the value of the \Lande\ factor obtained for a particular gate voltage $V_g$ can be considered as the maximum achievable value at that specific $V_g$.

\subsection{$\mathbf{k}\cdot\mathbf{p}$ theory of the \Lande\ factor}

Our model is based on the $8 \times 8 ~ \textbf{k} \cdot \textbf{p}$ approximation described by 
\begin{equation}
    \hat{\H}_{8 \times 8} = 
    \begin{bmatrix}
    \hat{\H}_c & \hat{\H}_{cv} \\
    \hat{\H}_{cv}^\dagger & \hat{\H}_v
    \end{bmatrix},
    \label{eq:8x8model}
\end{equation}
where $\hat{\H}_c$ is the Hamiltonian of the conduction band electrons corresponding to the $\Gamma_{6c}$ band. In the presence of the magnetic field $\hat{\H}_c$ can be written as
\begin{equation}
    \hat{\H}_c = H_{\Gamma_{6c}} \textbf{I}_{2 \times 2}+\frac{1}{2}\mu_Bg_0 \boldsymbol{\sigma}\cdot\mathbf{B},
    \label{Hc}
\end{equation}
where the second term corresponds to the Zeeman spin effect, $\mu_B$ is the Bohr magneton, $g_0$ is the \Lande\ factor of the free electron and $\boldsymbol{\sigma}=(\sigma_x,\sigma_y, \sigma_z)$ is the vector of Pauli matrices, while
\begin{equation}
    \hat{H}_{\Gamma_{6c}} = \frac{\hat{\textbf{P}}^2}{2m_0} + E_c + V(\mathbf{r}),
    \label{eq:gamma6c}
\end{equation}
where $\hat{\textbf{P}} = \hat{\textbf{p}} - e\textbf{A}$, $e$ is the electron charge, $m_0$~is the free electron mass and $E_c$ is the conduction band minima. The potential $V(\mathbf{r})$ in (\ref{eq:gamma6c}) contains the interaction of electrons with the electric field generated by the external gates $V_g(\mathbf{r})$ and the electron-electron interaction included in our model at the mean field level (Hartree potential) $V_H(\mathbf{r})$, $V(\mathbf{r})=V_g(\mathbf{r})+V_H(\mathbf{r})$.

Below we shall use a folding procedure of $\hat{\H}_{8 \times 8}$ to the conduction band sector, where in the Hamiltonian $\hat{\H}_v$, related to valance bands $\Gamma_{8v}$ and $\Gamma_{7v}$, all off-diagonal elements are neglected. Then, $\hat{\H}_v$ can be written as
\begin{equation}
    \hat{\H}_v = H_{\Gamma_{8v}} \textbf{I}_{4 \times 4} \oplus H_{\Gamma_{7v}}  \textbf{I}_{2 \times 2},
    \label{Hv}
\end{equation}
with
\begin{equation}
\begin{split}
    H_{\Gamma_{7v}} = E_{v'} & = E_c + V(\mathbf{r}) - E_0 - \Delta_0 \,,\\
    H_{\Gamma_{8v}} = E_v    & = E_c + V(\mathbf{r}) - E_0 \,,
\end{split}
\end{equation}
where $E_0$ is the energy gap and $\Delta_0$ is the energy of SO splitting in the valence band. Note that Eq.~(\ref{Hv}) neglects the kinetic term and Zeeman splitting in the valance band as the corresponding  energies are much smaller than $E_0$ and $\Delta_0$.

The coupling between the conduction band and the valence band is described by the off-diagonal matrix $\hat{\H}_{cv}$,
\begin{equation}
    \hat{\H}_{cv} = \frac{P_0}{\hbar}
    \begin{bmatrix}
    \frac{\hat{P}_+}{\sqrt{6}} & 0 &\frac{\hat{P}_-}{\sqrt{2}} & -\frac{\sqrt{2}\hat{P}_z}{\sqrt{3}}  & -\frac{\hat{P}_z}{\sqrt{3}} & \frac{\hat{P}_+}{\sqrt{3}}  \\ \\ 
    -\frac{\sqrt{2}\hat{P}_z }{\sqrt{3}}  & -\frac{ \hat{P}_+}{\sqrt{2}} & 0 & -\frac{\hat{P}_-}{\sqrt{6}} & \frac{\hat{P}_-}{\sqrt{3}} & \frac{\hat{P}_z }{\sqrt{3}} 
\end{bmatrix}\,,
\label{Hcv}
\end{equation}
where $\hat{P}_{\pm} = \hat{P}_x \pm i \hat{P}_y$ and the parameter $P_0 = \frac{-i\hbar}{m_0} \langle S|\hat{p}_x|X\rangle$ accounts for the coupling between conduction and valence bands at the $\Gamma$ point of the Brillouin zone. 

Using the standard folding-down transformation, we can reduce the $8\times 8$ $\mathbf{k}\cdot\mathbf{p}$ model (\ref{eq:8x8model}) into the effective $2\times 2$ Hamiltonian for  conduction electrons 
\begin{equation}
    \hat{\H}_{\mathit{eff}} = \hat{\H}_c - \hat{\H}_{cv}(\hat{\H}_v-E)^{-1}\hat{\H}_{cv}^\dagger = \hat{\H}_c + \Tilde{\H}_c.
    \label{Heff}
\end{equation}
In the above formula,  $\Tilde{\H}_c$ can be written in terms of Pauli matrices 
\begin{equation}
    \Tilde{\H}_c = \lambda_0\textbf{I}_{2 \times 2} + \boldsymbol{\lambda} \cdot \boldsymbol{\sigma},
    \label{Hcc}
\end{equation}
where
\begin{subequations}
\begin{align}
    \lambda_0 & = \frac{P_0^2}{3\hbar^2} \bigg [ \hat{P}_x \left (\frac{2}{E_v-E} + \frac{1}{E_{v'}-E}\right )\hat{P}_x  \nonumber \\
    & \:\:\:\:\:\:\:\:\:\:\:\:\:\:\:\:\:\:\:\:\:\:\:\:\ +\hat{P}_y \left (\frac{2}{E_v-E} + \frac{1}{E_{v'}-E}\right )\hat{P}_y  \bigg ],  
\label{alpha_0} \\
    \lambda_x & = \frac{iP_0^2}{3\hbar^2} \bigg [\hat{P}_z \left (\frac{1}{E_v-E} - \frac{1}{E_{v'}-E} \right)\hat{P}_y \nonumber \\ 
    & \:\:\:\:\:\:\:\:\:\:\:\:\:\:\:\:\:\:\:\:\:\:\:\:\
    - \hat{P}_y\left(\frac{1}{E_v-E} - \frac{1}{E_{v'}-E}\right )\hat{P}_z \bigg ],    
\label{alpha_x} \\
    \lambda_y & = \frac{iP_0^2}{3\hbar^2}\bigg[ \hat{P}_x\left(\frac{1}{E_v-E} - \frac{1}{E_{v'}-E}\right)\hat{P}_z \nonumber \\
    & \:\:\:\:\:\:\:\:\:\:\:\:\:\:\:\:\:\:\:\:\:\:\:\:\
    - \hat{P}_z\left(\frac{1}{E_v-E} - \frac{1}{E_{v'}-E}\right)\hat{P}_x \bigg ], 
\label{alpha_y} \\
    \lambda_z & = \frac{iP_0^2}{3\hbar^2}\bigg[\hat{P}_y\left (\frac{1}{E_v-E} - \frac{1}{E_{v'}-E}\right )\hat{P}_x \nonumber \\
    & \:\:\:\:\:\:\:\:\:\:\:\:\:\:\:\:\:\:\:\:\:\:\:\
    - \hat{P}_x\left(\frac{1}{E_v-E} - \frac{1}{E_{v'}-E}\right )\hat{P}_y \bigg ].        
\label{alpha_z}
\end{align}
\end{subequations}
The first term in Eq.~(\ref{Hcc}) leads to the standard formula for the effective mass 
\begin{equation}
\begin{aligned}
    \frac{1}{m^*} = \frac{1}{m_0}+\frac{2P_0^2}{3\hbar^2} \left ( \frac{2}{E_v} + \frac{1}{E_{v'}} \right )\,,
\end{aligned}
\end{equation}
while the second term corresponds to the Rashba SO coupling. If we assume that $E_0$ and $\Delta_0$ are the largest energies in the system 
we can expand $E_{v(v')}$ in Eqs.~(\ref{alpha_x}-\ref{alpha_z}) to the second order in energy. Then, Eqs.~(\ref{alpha_x})-(\ref{alpha_z}) can be rewritten as
\begin{subequations}
\begin{align}
    \label{lx}
    \lambda_x & = - \alpha_R^y \left( k_z - \frac{e}{\hbar}A_z \right ) - \frac{eP_0^2}{3\hbar}\left ( \frac{1}{E_0} - \frac{1}{E_0 + \Delta_0} \right )B_x,\\
    \label{ly}
    \lambda_y & = \alpha_R^x \left( k_z - \frac{e}{\hbar}A_z \right ) - \frac{eP_0^2}{3\hbar}\left ( \frac{1}{E_0} - \frac{1}{E_0 + \Delta_0} \right )B_y, \\
    \label{lz}
    \lambda_z & = \alpha_R^y \left( \hat{k}_x - \frac{e}{\hbar}A_x \right ) - \alpha_R^x \left( \hat{k}_y - \frac{e}{\hbar}A_y \right ) \nonumber \\ 
    & \:\:\:\:\:\:\:\:\:\:\:\:\:\:\: - \frac{eP_0^2}{3\hbar}\left ( \frac{1}{E_0} - \frac{1}{E_0 + \Delta_0} \right )B_z,
\end{align}
\end{subequations}
where
\begin{equation}
\begin{split}
    \boldsymbol{\alpha}_R & = (\alpha_R^x, \alpha_R^y, \alpha_R^z) \\
     & = \frac{P_0^2}{3} \left ( \frac{1}{E_0^2} - \frac{1}{\left ( E_0 + \Delta_0 \right )^2} \right )\boldsymbol{\nabla} V(x,y)
\end{split}
\label{alpha_rashba}
\end{equation}
is the Rashba SO coupling constant and we assume $\hat{p}=\hbar (\hat{k}_x, \hat{k}_y, k_z)=\hbar(-i \partial / \partial x, -i \partial / \partial y, k_z)$. Note that in Eqs.~(\ref{lx}, \ref{ly}) we have already omitted $\alpha_R^z$ terms since the magnetic field does not break translational invariance along the wire axis, i.e., 
\begin{equation}
\begin{split}
    \Psi_{n,k_z}(x,y,z) & =\psi_{n,k_z}(x,y)e^{ik_zz} \\
    & =[\psi_{n,k_z}^\uparrow(x,y),\psi_{n,k_z}^\downarrow(x,y)]^Te^{ik_zz}\,.
\end{split}    
\end{equation}

From $\hat{\H}_{\mathit{eff}}$ we determine the spin-split energy subbands $E_{n,k_z}(\mathbf{B})$, and from these the effective $\gstar$ factor of the lowest state as
\begin{equation}
    \gstar=\frac{(E_{2,kz}(\mathbf{B})-E_{2,kz}(0))-(E_{1,kz}(\mathbf{B})-E_{1,kz}(0) ) }{\mu_b \hbar B}.\
    \label{eq:g_energy}
\end{equation}
Note that the above definion of $\gstar$ excludes the spin splitting which is due to the SO coupling solely, and may be present also at $B=0$ (at which $\gstar=0$). However, the total SO term involves the magnetic field by the kinetic momentum, and it also contributes to the effective \Lande\ factor.
To show that, let us decompose the SO term into the part depending on the canonical momentum $\mathbf{k}$ and the vector potential, $\mathbf{A}$. Then, the effective Hamiltonian for conduction electrons can be written as
\begin{eqnarray}
    \hat{\H}_{\mathit{eff}}&=&\left ( \frac{\hat{\textbf{P}}^2}{2m^*} + E_c + V(\mathbf{r}) \right ) \textbf{I}_{2 \times 2}+(\alpha_R^x\sigma_y-\alpha_R^y\sigma_x)k_z \nonumber \\
    &+& (\alpha_R^y \hat{k}_x-\alpha_R^x \hat{k}_y)\sigma_z + \frac{1}{2}\mu_B \mathbf{B} \tgstar \boldsymbol{\sigma}
    \label{full_hamiltonian}
\end{eqnarray}
where $\tgstar$ is a tensor given by
\begin{equation}
    \tgstar  = g_{RLZ}\mathbf{I}_{3\times3}+\mathbf{g}_{SO}\,,
    \label{gfull}
\end{equation}
and  
\begin{equation}
g_{RLZ} = g_0 - \frac{2E_p}{3} \left(\frac{1}{E_0} -\frac{1}{E_0 + \Delta_0} \right)\,,
\end{equation}
which corresponds to the well-know \RLZ\ formula,\cite{Roth-Lax} ($E_p = {2m_0P_0^2}/{\hbar^2}$), while the tensor $\mathbf{g}_{SO}$ results from the orbital effects of the magnetic field in the SO Hamiltonian,
\begin{equation}
    \mathbf{g}_{SO}  = 
    \begin{bmatrix}
        g^{xx}_{SO} & g^{xy}_{SO} & 0 \\
        g^{yx}_{SO} & g^{yy}_{SO} & 0 \\
        0 & 0 &  g^{zz}_{SO}
    \end{bmatrix}.
    \label{g_tensor}
\end{equation}

Using the vector potential (\ref{eq:vec_pot}), the elements of this tensor can be expressed as
\begin{subequations}
\label{g_ab}
\begin{align}
   g^{xx}_{SO} & =  \frac{2e}{\mu_B\hbar} \alpha_R^y y, \label{gxx} \\
   g^{yy}_{SO} & =  \frac{2e}{\mu_B\hbar} \alpha_R^x x, \label{gyy}\\
   g^{zz}_{SO} & = \frac{e}{\mu_B\hbar} \left( \alpha_R^y y - \alpha_R^x x \right ), \label{gzz} \\
   g^{xy}_{SO} & = -\frac{2e}{\mu_B \hbar}\alpha_R^x y, \label{gxy} \\
   g^{yx}_{SO} & = -\frac{2e}{\mu_B \hbar}\alpha_R^y x\label{gyx}.
\end{align}
\end{subequations}
which shows that $\tgstar$ depends linearly on the vector of Rashba SO coupling constants $\boldsymbol{\alpha}_R$. 

Note that $\mathbf{g}_{SO}$ is not an observable and it is gauge dependent (while of course $\lambda_x(y,z)$ in Eqs.~(13a-13c), hence $\gstar$, are gauge invariant, as they involve the
kinetic momentum $P$). However, since it explicitly demonstrates the  contribution to the effective \Lande\ factor from the SO coupling, it is useful to use $\mathbf{g}_{SO}$ for analysing $\gstar$.

Since the Rashba coefficients and the SO induced \Lande\ factor are functions of space [see Eqs.~(\ref{alpha_rashba}, \ref{g_ab})], we discuss the matrix elements of the Rashba SO coupling constants
\begin{equation}
       \langle {\alpha}{_{R}^{x(y)}}(k_z) \rangle _n = \langle \psi_{n,k_z} |\alpha_R^{x(y)} \sigma_{y(x)}|\psi_{n,k_z} \rangle
\label{eq:el_alfa}
\end{equation}
and the individual diagonal and off-diagonal matrix elements of $\mathbf{g}_{SO}$, respectively defined as
\begin{subequations}
\label{eq:el_g}
\begin{align}
   \langle g_{SO}^{xx(yy,zz)}(k_z) \rangle _n &= \langle \psi_{n,k_z} |g^{xx(yy,zz)}_{SO}\sigma_{x(y,z)}|\psi_{n,k_z} \rangle, \label{eq:el_gdiag} \\
       \langle g_{SO}^{xy(yx)}(k_z) \rangle _n &= \langle \psi_{n,k_z} |g^{xy(yx)}_{SO} \sigma_{y(x)}|\psi_{n,k_z} \rangle,
\end{align}
\end{subequations}
where $|\psi_{n,k_z} \rangle$ is the in-plane part of the $n$-th envelope functions of NW, to be calculated as described in the following section.

It is useful to compare our derivation with Lassnig's for the two dimensional gas, reported in Ref.~\onlinecite{Lassnig}. There, $\gstar$ has been defined in such a way that its first derivative determines the SO coupling constant, hence it contains information about the \textit{total} spin splitting of the energy levels, ensuing both from the linear Zeeman term and the SO coupling, whose dependence on the magnetic field is more complex. Here, instead, we define the effective \Lande\ factor as the coefficient of proportionality between the spin splitting of the energy levels induced by the external magnetic field and the magnitude of the field.  This procedure allows to distinguish between two effects among which the one which changes with $\mathbf{B}$ defines the effective \Lande\ factor. Note that such a definition is usually used in experiments to determine $\gstar$.\cite{Schroer,vanWeperen2013,Marcus2018}

\subsection{Numerical calculations}

To understand the physics behind the behaviour of the \Lande\ factor in NWs with strong SO coupling, we use a numerical approach taking into account important key ingredients, namely the orbital and Zeeman effect, SO coupling and electrostatic environment. For this purpose, we employ a standard Shr{\"o}dinger-Poisson approach.\cite{Wojcik2018, Wojcik2021,Bertoni2011,Vezzosi2022,Woods2018,Stern1972,Ando1976} Assuming the translational invariance along the growth axis $z$, the envelope functions $\psi_{n,k_z}(x,y)=[\psi_{n,k_z}^\uparrow(x,y),\psi_{n,k_z}^\downarrow(x,y)]$ can be determined from the Schr{\"o}dinger equation 
\begin{eqnarray}
    &\bigg [& \bigg ( \frac{\boldsymbol{\hat{P}}_{2D}^2}{2m^*} + \frac{1}{2}m^*\omega_c^2[(y\cos\theta-x\sin\theta)\sin\varphi-k_zl_B^2]^2 + E_c \nonumber \\
    &+& V(\mathbf{r}) \bigg )  \textbf{I}_{2 \times 2}+(\alpha_R^x\sigma_y-\alpha_R^y\sigma_x)k_z 
    + (\alpha_R^y \hat{k}_x-\alpha_R^x \hat{k}_y)\sigma_z \nonumber \\
    &+& \frac{1}{2}\mu_B \mathbf{B} \tgstar \boldsymbol{\sigma} \bigg ] \psi_{n,k_z}(x,y)=E_{n,k_z}\psi_{n,k_z}(x,y),
    \label{Hc2D}
\end{eqnarray}
where $\alpha_R^{x(y)}$ and $\tgstar$ are functions of the position $(x,y)$, $\omega_c=eB/m^*$ is the cyclotron frequence, $l_B=\sqrt{\hbar/eB}$ is the magnetic length and
\begin{equation}
    \hat{\boldsymbol{P}}_{2D}^2= \left  ( \hat{p}_x+eB\frac{y\cos \varphi}{2} \right )^2 + \left  ( \hat{p}_y-eB\frac{x\cos \varphi}{2} \right )^2.
\end{equation}
Note that in the presence of magnetic field and SO coupling the Hamiltonian (\ref{Hc2D}) depends on the $k_z$ vector. The calculations are carried out on a uniform grid in the range $[-k_z^{max},k_z^{max}]$ where $k_z^{max}$ is chosen to be much larger than the Fermi wave vector. The term $(\alpha_R^y \hat{k}_x-\alpha_R^x \hat{k}_y)\sigma_z$ in Hamilonian (\ref{Hc2D}) needs an additional comment as it may suggest the violation of time reversal symmetry. As we checked, this is not the case and $[(\alpha_R^y \hat{k}_x-\alpha_R^x \hat{k}_y)\sigma_z,\mathcal{T}]=0$, where $\mathcal{T}=\mathcal{K}(-y\sigma_y)$ and  $\mathcal{K}$ is the complex conjugate operator. As a result, at $\mathbf{B}=0$ the Kramers degeneracy is preserved, resulting in the crossing of states at $k_z=0$.

The self-consistent potential $V(\mathbf{r})$ in Eq.~(\ref{Hc2D}) is determined at the mean field level by solving of the Poisson equation 
\begin{equation}
    \nabla _{2D}^2V(x,y)=-\frac{n_e(x,y)}{\epsilon_0\epsilon}
    \label{Poisson}
\end{equation}
where $\epsilon$ is a dielectric constant and the electron density $n_e$ can be calculated based on the formula
\begin{equation}
    n_e(x,y)=\sum_n \int_{-k_z^{max}}^{k_z^{max}}\frac{1}{2\pi} |\psi_{n,k_z}(x,y)|^2 f(E_{n,k_z}-\mu,T) dk_z
\end{equation}
where $\mu$ is the chemical potential,  $T$ is the temperature and $f(E,T)$ is the Fermi-Dirac distribution.

In the applied Shr{\"o}dinger-Poisson approach, equations (\ref{Hc2D}) and (\ref{Poisson}) are solved alternatively until the self-consistency is reached, which we consider to occur when the relative variation of the charge density between two consecutive iterations is lower than $0.001$. In each iteration a spatial distribution of $\alpha_R^{x(y)}$ and $g_{SO}^{ab}$, where $a,b=\{x,y,z\}$, are determined based on Eqs.~(\ref{alpha_rashba}) and (\ref{g_ab}). Numerical calculations are carried on the triangular grid, which preserves the hexagonal symmetry of the Hamiltonian at zero field, avoiding artifacts such as spurious level splittings which may appear when using rectangular grid symmetry.\cite{Bertoni2011} We assume Dirichlet boundary condition for all the facets with a specified condition for the bottom one, defined by the voltage applied to the gate. Finally, the energy spectrum $E_{n,k_z}$, the self-consistent potential $V(x,y)$ and the corresponding wave functions $\psi_{n,k_z}(x,y)$ are used to determine $\gstar$, $\langle \alpha_R^{x,(y)} \rangle _n$ as well as $\langle g^{{xx(yy,zz)}}_{SO} \rangle _n$ and $\langle g^{{xy(yx)}}_{SO} \rangle _n$ tensor elements according to Eqs.~(\ref{eq:g_energy}, \ref{eq:el_alfa}, \ref{eq:el_g}).

Calculations have been carried out for the material parameters corresponding to InSb: $E_0=0.235$~eV, $\Delta_0=0.81$~eV, $m^*=0.014$, $E_P= \frac{2m_0P}{2} = 23.3$~eV, $T=4.2$~K, and for the nanowire width $W=100$~nm (corner-to-corner). We keep the constant linear electron density at the low level $n_e=8\times 10^7$ cm$^{-1}$ which guarantees that only the lowest subband is occupied in the range of the considered magnetic field $B=[0,4]$~T.

\section{Results}

We shall now discuss the effective \Lande\ factor as a function of the magnetic field intensity and direction. As $g_{RLZ}$ evaluated from the \RLZ\ formula ($g_{RLZ}=-49$ for the present material) does not depend on the magnetic field, we put particular emphasis on the role of the SO-induced component $\mathbf{g}_{SO}$ in terms of the tensor elements, Eqs.~(\ref{g_ab}). As shown in the previous section, corrections to the \Lande\ factor coming from the SO interaction are indirectly dependent on the wave-vector via $\psi_{n,k_z}$, which results from  the orbital effects of the magnetic field. For this reason, we shall study both $\gstar$ and $\mathbf{g}_{SO}$ as a function of both the wave vector and the magnetic field. We limit our study to the lowest subband assuming the electrical potential is applied to the bottom gate to induce SO coupling. For simplicity, in the rest of the paper we omit the subband index in Eqs.~(\ref{eq:el_alfa}), (\ref{eq:el_g}), i.e. $\langle \dots \rangle _{n=1}=\langle \dots \rangle$.

\subsection{Enhancement of the \Lande\ factor due to SO coupling}

First, we show that a magnetic field oriented along the $x$ axis, i.e., perpendicular to the NW axis \emph{and} to the direction of $\langle\boldsymbol{\alpha}_R\rangle$, results in a substantial enhancement of the effective \Lande\ factor. For this purpose, we assume that $V_g=0.2$~V is applied to the bottom gate, generating an electric field that mantains reflection symmetry with respect to the $y$ axis; hence $\langle\boldsymbol{\alpha}_R\rangle$ is directed along $y$ by symmetry.
\begin{figure}[!htp]
    \centering
    \includegraphics[width = 0.5 \textwidth]{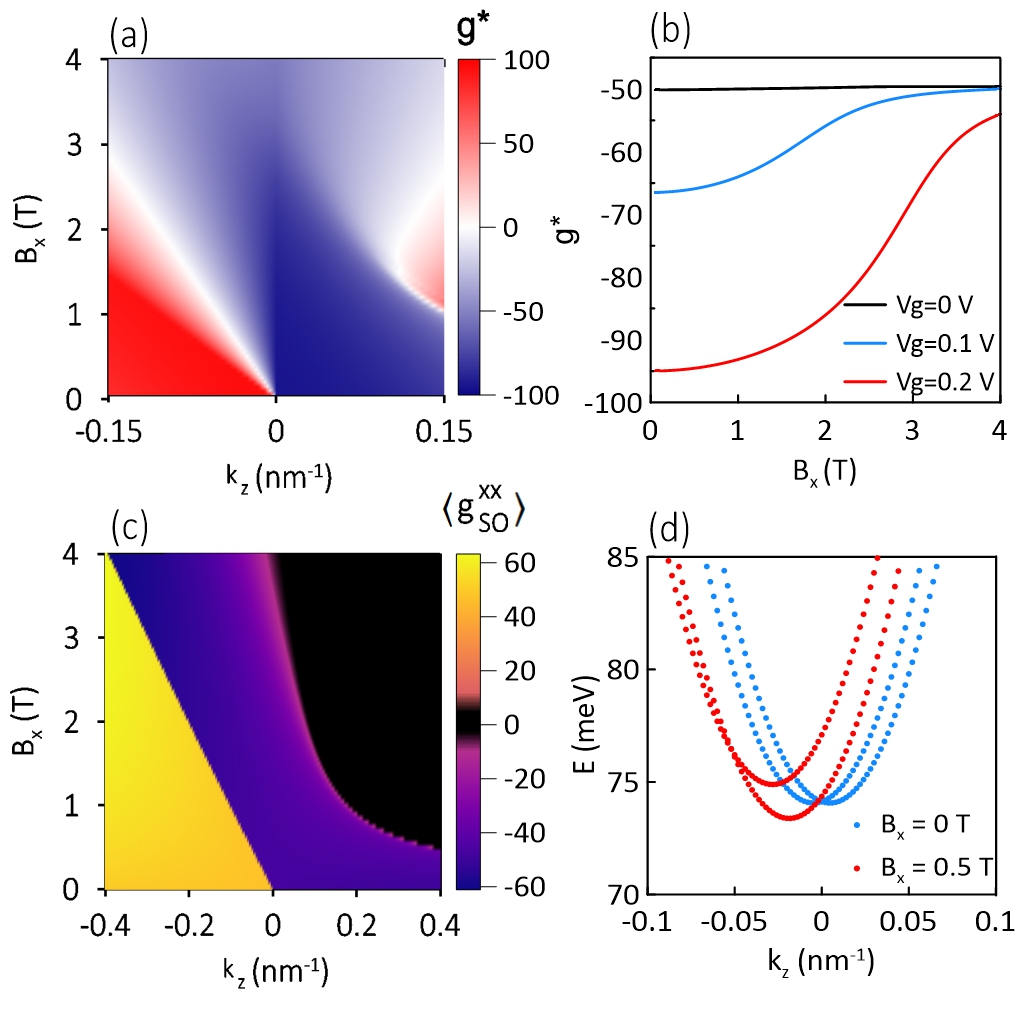}
\caption{ (a) Map of $\gstar$ [Eq.~(\ref{eq:g_energy})] as a function of wave vector $k_z$ and magnetic field oriented along the $x$-axis, $B_x$. (b) $\gstar (B_x)$ calculated at $k_z=0$ at selected $V_g$. (c) Map of $\langle g_{SO}^{xx} \rangle$ as a function of wave vector $k_z$ and magnetic field, $B_x$. (b) Dispersion relations without magnetic field (blue) and at $B_x = 0.5$~T (red). The shift of the crossing point on the panel (d) corresponds to the sign change of $\langle g_{SO}^{xx} \rangle$ in panel (c).}
    \label{fig:2}
\end{figure}

In Fig.~\ref{fig:2}(a) we show the effective \Lande\ factor $\gstar$ [see Eq.~(\ref{eq:g_energy})] vs $k_z$ and $B$. In this configuration $|\gstar|$ reaches values up to $100$,  twice as large as  predicted from the RLZ formula ($g_{RLZ}=-49$). The maximum of $|\gstar|$ is determined by the gate voltage, as shown in Fig.~\ref{fig:2}(b) where we report the calculated $\gstar(B_x)$ at $k_z=0$ for selected values of $V_g$. Note that at $V_g=0$, when the SO coupling is absent, $\gstar=g_{RLZ}$ which strongly suggests that the observed enhancement of the \Lande\ factor is related to the orbital effects in the SO term. In order to show that, in Fig.~\ref{fig:2}(c), we present the map of the diagonal element  $\langle g_{SO}^{xx}\rangle(k_z,B_x)$. Note that with this field configuration the off-diagonal elements vanish by symmetry. Indeed, the reflection symmetry of the electric field with respect to the $y$ axis leads to $\langle \alpha_R^x \rangle =0$, hence $\langle g_{SO}^{xy} \rangle =0$ [see Eq.~(\ref{gxy})]. Moreover, the even symmetry of the envelope function is unaffected by the magnetic field directed along $x$,  hence $\langle g_{SO}^{yx} \rangle=0$ [see Eqs.~(\ref{gyx})]. Fig.~\ref{fig:2}(c) clearly demonstrates that the correction to the effective \Lande\ factor arising from the orbital effects in the SO coupling term reaches a value similar to that obtained from the RLZ formula. Under certain conditions, this enhancement can lead to a significant increase of $\tgstar$, almost doubling it, as observed in recent experiments.~\cite{Schroer,vanWeperen2013,Marcus2018}

In Fig.~\ref{fig:2}(c) we distinguish three regions, with positive (yellow), negative (purple) and vanishing (black) $\langle g_{SO}^{xx} \rangle$. The abrupt change of sign between positive and negative regions is simply understood as the crossing of subbands of opposite spin, since only the value for the lowest subband is shown here. Indeed, as shown in Fig.~\ref{fig:2}(d), the subband of opposite spin cross at $k_z=0$ at vanishing field. When the field is switched on, both subband shift to negative $k_z$ and shift in energy due to Zeeman term. Hence, the crossing shifts linearly with the field to more negative wavevectors, as shown in  Fig.~\ref{fig:2}(c).
\begin{figure}[!htp]
    \centering
    \includegraphics[width = 0.4 \textwidth]{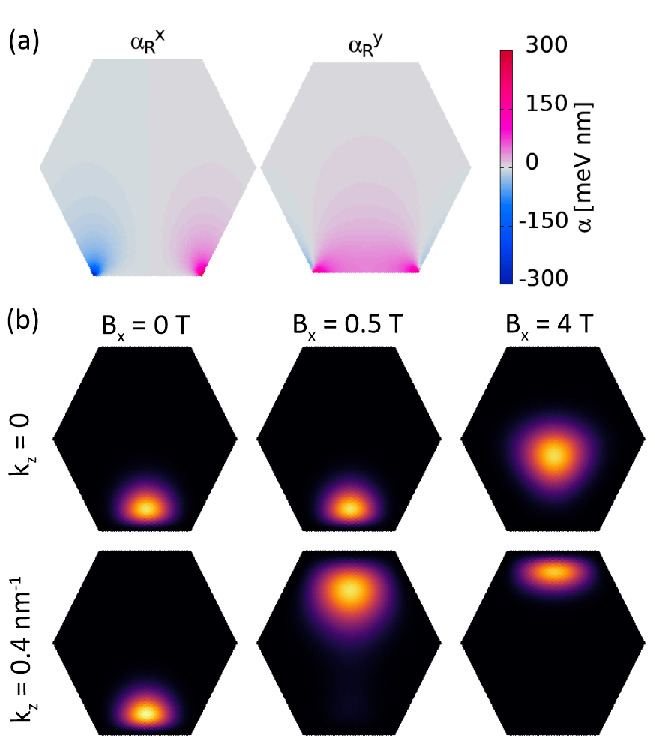}
    \caption{(a) Map of the SO Rashba coefficients $\alpha_R^{x}, \alpha_R^{y}$. (b) Squared envelope functions of the lowest subband with the magnetic field oriented along $x$, at selected magnetic field intensity $B_x$ and the wave vectors $k_z$.}
    \label{fig:alpha_psi}
\end{figure}

For sufficiently large $k_z>0$ and field intensity, $\langle g_{SO}^{xx} \rangle$ almost vanishes, as shown in Fig.~\ref{fig:2}(c) - black region. This can be explained by the analysis of the position-dependent SO coupling constants $\alpha_R^{x(y)}$ [see Eq.~(\ref{alpha_rashba})] presented in the Fig.~\ref{fig:alpha_psi}(a) at $B=0$. Note that their spatial distribution is primarily influenced by the electric field generated by the bottom gate and do not undergo significant changes as the magnetic field increases. 
Since the value of $\mathbf{g}_{SO}$ matrix elements depends on the Rashba SO coupling constant, the SO-induced modification of the \Lande\ factor for a specific subband is most significant when its envelope function is localized in the regions of strong Rashba SO coupling. With this respect, the vanishing of $\langle g_{SO}^{xx} \rangle$ in Fig.~\ref{fig:2}(c) is due to the change of the wave function localization, determined by the orbital coupling to the magnetic field.

\begin{figure}[!htp]
    \centering
    \includegraphics[width = 0.5 \textwidth]{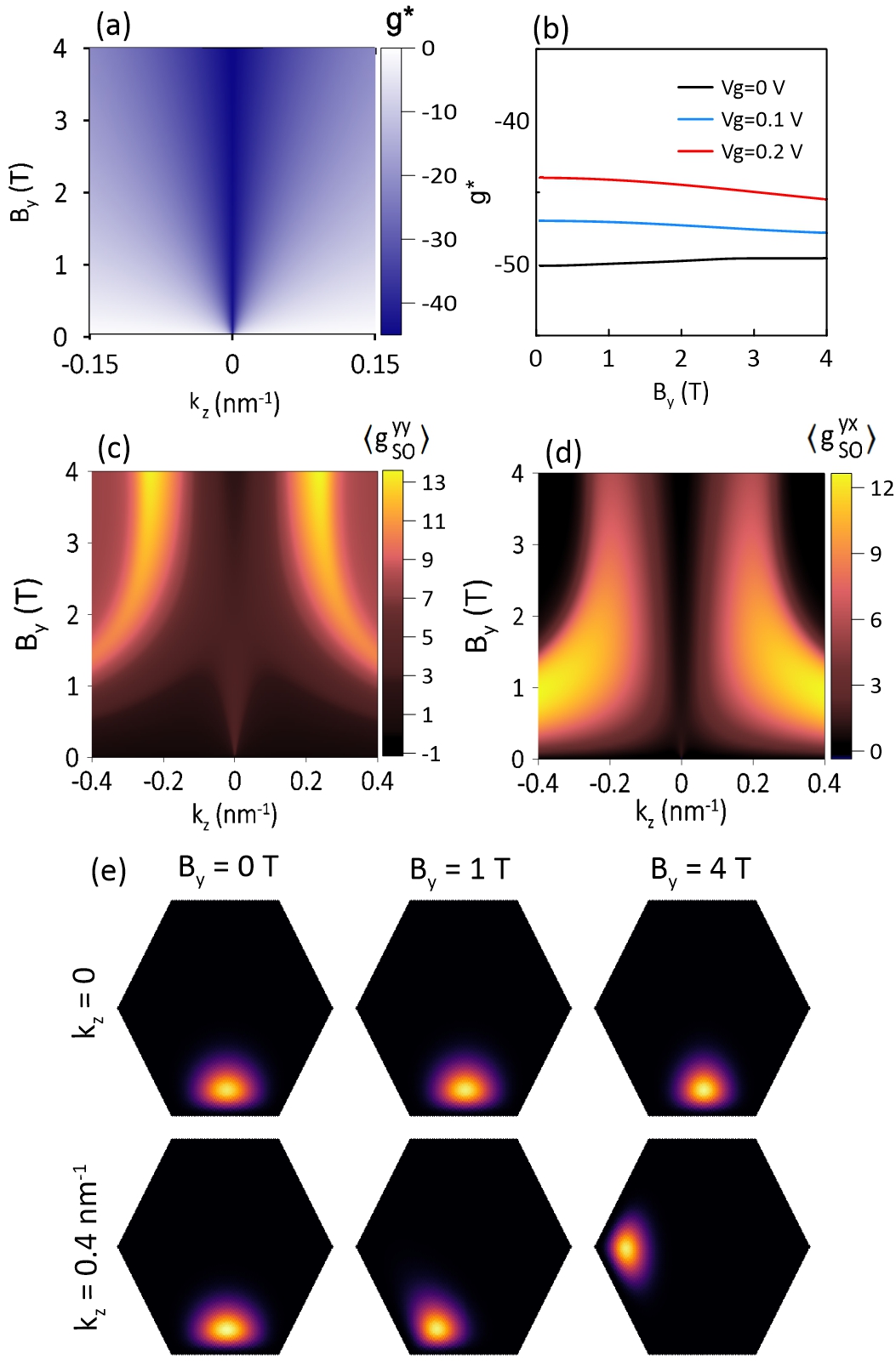}
\caption{(a) Map of $\gstar$ [Eq.~(\ref{eq:g_energy})] as a function of wave vector $k_z$ and magnetic field oriented along the $y$-axis, $B_y$. (b) $\gstar (B_y)$ calculated at $k_z=0$. (c,d) Diagonal $\langle g_{SO}^{yy} \rangle$ and off-diagonal $\langle g_{SO}^{yx} \rangle$ elements of $\mathbf{g}_{SO}$ with the field oriented along $y$, as a function of wave vector $k_z$ and field intensity $B_y$. (e) Squared envelope functions of the lowest energy state at selected different magnetic field intensity $B_y$ and the wave vectors $k_z$. }
    \label{fig:gyy_magnitude}
\end{figure}

In Fig.~\ref{fig:alpha_psi}(b), we report the squared envelope functions of the lowest subbands at $k_z=0$ and $k_z=0.4$ nm$^{-1}$ at increasing magnetic fields. At $k_z=0$ there is no kinetic coupling to the magnetic field and the localization of the envelope function is only determined by the electric field; hence, it  concentrates near the bottom gate, where the SO coupling is strong. For a positive wave vectors $k_z$, instead, the orbital effects shift the wave function towards the opposite facet of the NW, where the SO coupling is weak, leading to  vanishing  $\langle g^{xx}_{SO} \rangle$, which explains the black region in Fig.~\ref{fig:2}(c). As shown in Fig.~\ref{fig:2}(c), the stronger the magnetic field, the lower $k_z$ is required to push the wave function away from the region with large SO coupling, near the bottom facet. Naively, one might expect that the state $k_z=0$ would not be affected by this phenomenon as there is not orbital coupling to the magnetic field for this state. However, it should be noted that for high magnetic fields, diamagnetic effects become dominant, causing the wave functions to localize in the middle of NW along the field direction, resembling dispersionless Landau levels, as shown in Fig.~\ref{fig:alpha_psi}(b). As the position of this wave function is associated with low SO coupling regions, $\langle g^{xx}_{SO} \rangle$ gradually decreases towards zero, even for $k_z=0$, as illustrated in Fig.~\ref{fig:2}(c). Thus, regardless of the gate voltage, $\gstar$ tends to approach $g_{RLZ}$ when the magnetic field increases - see Fig.~\ref{fig:2}(b) .
\begin{figure}[!t]
    \centering
    \includegraphics[width = 0.3 \textwidth]{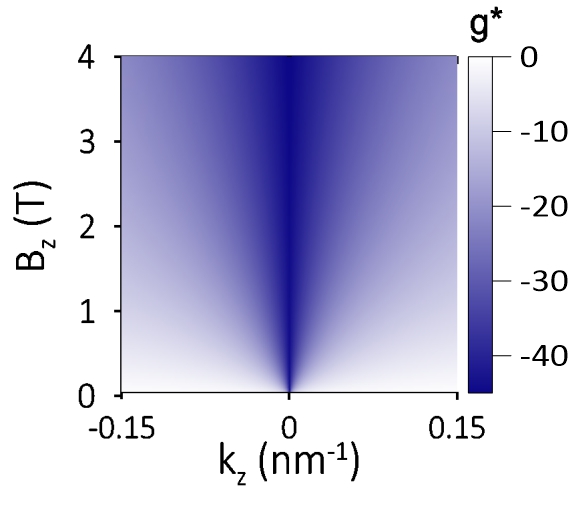}
    \caption{Effective \Lande\ factor $\gstar$ as a function of wave vector $k_z$ and magnetic field magnitude oriented in the $z$-direction, $B_z$. }
    \label{fig:gzz_magnitude}
\end{figure}

We next discuss the behavior of the effective \Lande\ factor with the magnetic field directed either parallel to $\boldsymbol{\alpha}_R$ (along the $y$ axis) or to the NW axis (along the $z$ axis). When the magnetic field is applied parallel to $\boldsymbol{\alpha}_R$, $\langle g^{yy(yx)}_{SO} \rangle \ge 0$, resulting in the increase of $\gstar$. This is  shown in Fig.~\ref{fig:gyy_magnitude}(a,b). In this case, the deviation from $g_{RLZ}$ is not as large as for the perpendicular orientation of $\mathbf{B}$ -- compare with Fig.~\ref{fig:2}(a).  
In this configuration the off-diagonal element $\langle g_{SO}^{yx} \rangle $ is non-negligible, in contrast to $\langle g_{SO}^{xy} \rangle$ which is nearly zero, as the avarege value of $\alpha_R^x$ is vanishing due to the gate symmetry. Again, the evolution of both $\langle g_{SO}^{yy} \rangle$ and $\langle g_{SO}^{yx} \rangle$ as a function of the magnetic field, shown in Fig.~\ref{fig:gyy_magnitude}(c) and Fig.~\ref{fig:gyy_magnitude}(d), respectively, is determined by the localization and symmetry of the wave function, whereas we assume the rule that we display only these tensor elements which contribute to the spin splitting for a particular field direction.

In Fig.~\ref{fig:gyy_magnitude}(e), one can observe that at zero magnetic field, the wave function sets itself at the center-bottom of the NW. In this region, $\alpha_R^x$ is antisymmetric with respect to the $x$ axis, resulting in the $\langle g_{SO}^{yy} \rangle=\langle g_{SO}^{yx} \rangle= 0$. The symmetry of the wave function is broken by the magnetic field, as depicted in Fig.~\ref{fig:gyy_magnitude}(e). For $k_z = 0.4$ nm$^{-1}$, for increasing magnetic fields, the wave function is first localized at the bottom-left corner, where the contribution from negative $\alpha_R^x$ leads to non-zero values of $\langle g_{SO}^{yy(yx)} \rangle$, and eventually in the left corner, where $\alpha_R^x$ is significantly lower, resulting in a decrease in $\langle g_{SO}^{yy(yx)} \rangle$. This field-induced evolution leads to the maximum of $\langle g_{SO}^{yy(yx)} \rangle$ at a certain $k_z$ value, as illustrated in Fig.~\ref{fig:gyy_magnitude}(c,d).

We next consider a magnetic field applied in $z$-direction, i.e., along the NW axis. Decrease of $|\gstar|$, shown in Fig.~\ref{fig:gzz_magnitude}, has a different nature, since the orbital effects of magnetic field are highly reduced by the confinement. In this case the  localization of the wavefunction is not measurably changed with the magnetic field, regardless of $k_z$, and thus it does not determine the evolution of $\gstar$ with $k_z$ and $B_z$. In this configuration $\gstar$ is rather governed by the interplay between the Zeeman effect, which favors in-wire $z$ polarization, and the SO interaction, which favors orthogonal polarization along $y$. Note that both the tensor element $\langle g_{SO}^{zz} \rangle $ [see Eqs.~(\ref{eq:el_g})] and the total $\gstar$ factor, are defined by the energy splitting which depends on $\sigma_z$ and thus to the relative distribution of spin up and down component in the spinor. Since the SO coupling depends on the wave vector, for a small $k_z$ the ordinary Zeeman effect is dominant, aligning the electron spin along the magnetic field direction and - in the limit of $k_z=0$ - makes the system spin polarized along the $z$ axis. The expectation value of $\sigma_z$ in this case is the largest in the sense of absolute value, resulting in the large value of $\gstar$. In other words, the value of $\gstar$ for small $k_z$ results from the finite Rashba couplings near the bottom gate, where the wave function is localized and the almost complete $z$-spin polarization of electrons induced by the magnetic field. As a consequence, $\gstar$ is independent of the magnetic field magnitude at $k_z=0$ (not shown here).
\begin{figure}[!htp]
    \centering
    \includegraphics[width = 0.48 \textwidth]{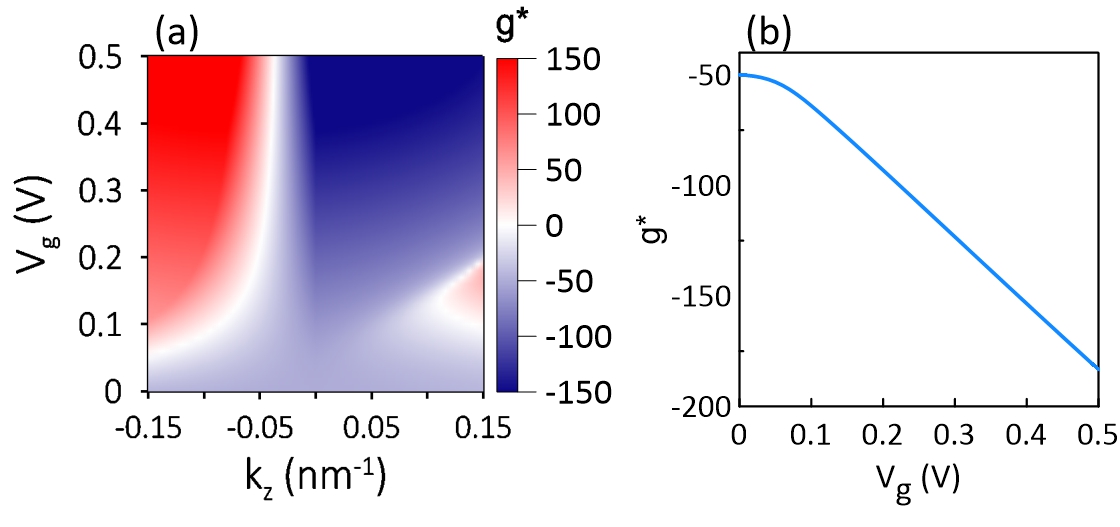}
    \caption{(a) $\gstar$ as a function of wavevector $k_z$ and bottom gate voltage $V_g$ and (b) $\gstar(V_g)$ at $k_z=0$. Results for magnetic field directed along the $x$ axis with $B_x=1$\,T.  }
    \label{fig:gxx_Vg}
\end{figure}

On the other hand, for a large value of $k_z$ and low magnetic field, the SO coupling plays a major role, forcing the electron spin to align along the effective Rashba field directed in the $x$ axis. In this scenario, the spin-up and spin-down components of the spinor become almost equal, resulting in a decrease in $\gstar$. It is worth noting that even for a large $k_z$ and strong SO coupling, an increasing magnetic field can deviate the electron spin direction from the $x$ towards the $z$ axis, leading to an overall increase in $\gstar$ with the magnetic field, as depicted in Fig.~\ref{fig:gzz_magnitude}. 

Finally, note that results presented in Fig.~\ref{fig:gzz_magnitude} for the magnetic field directed along the $z$-axis  at $k_z=0$ corresponds to the physical situation considered theoretically in Ref.~\onlinecite{Winkler2017}, where the enhancement of the effective \Lande\ factor has been recently predicted in semiconductor NWs. The predicted effect was however restricted to the higher subbands characterized by the nonzero orbital momentum where the orbital effects are relevant. Here, we show that the enhancement of $g^*$ for the lowest band is possible only when the magnetic field is applied perpendicular to $\boldsymbol{\alpha}_R$ - in our setup along the $x$-axis.

To summarize this section, in Fig.~\ref{fig:gxx_Vg} we show the gate voltage dependence of $\gstar$, calculated for a magnetic field directed along the $x$ axis with $B_x = 1$~T. It can be observed that the inclusion of the SO effects may lead to a substantial increase of the effective \Lande\ factor $\gstar$, reaching up to four times the value obtained from the RLZ formula.

\subsection{Spin-orbital induced \Lande\ factor anisotropy}

We next analyze the anisotropy of $\gstar$ with respect to the field direction. For this purpose we consider a magnetic field with intensity $B=1$\,T rotated in (i) the $xz$ plane ($\varphi = 0$), (ii) the $xy$ plane ($\theta = \pi/2$) and (iii) the $yz$ plane ($\varphi = \pi /2$). To induce Rashba SO coupling we apply a gate voltage $V_g = 0.2$~V.
\begin{figure}[!htp]
    \centering
    \includegraphics[width = 0.4 \textwidth]{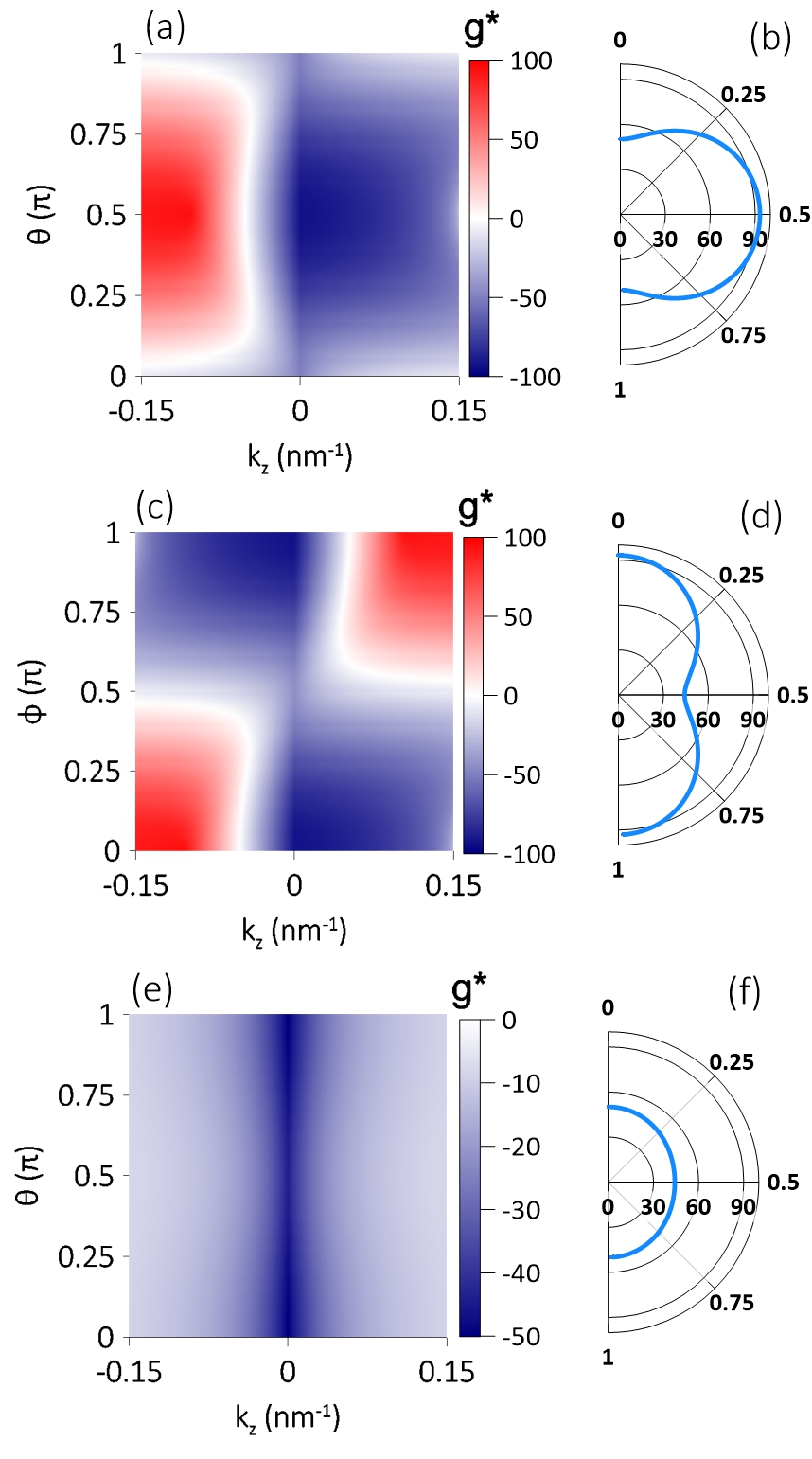}
    \caption{Maps of $\gstar$ [Eq.~(\ref{eq:g_energy})] as a function of wave vector $k_z$ and magnetic field orientation when it is rotated
    (a), (b) in the $xz$ plane;  (c), (d) in the $xy$ plane; (e), (f) in the $yz$ plane. Right polar plots present $\gstar$ evaluated at $k_z=0$. Results for $B=1$~T and $V_g=0.2$~V.}  
    \label{fig:Lande_anisotropy_gstar}
\end{figure}

\begin{figure*}[!htp]
    \centering
    \includegraphics[width = 1 \textwidth]{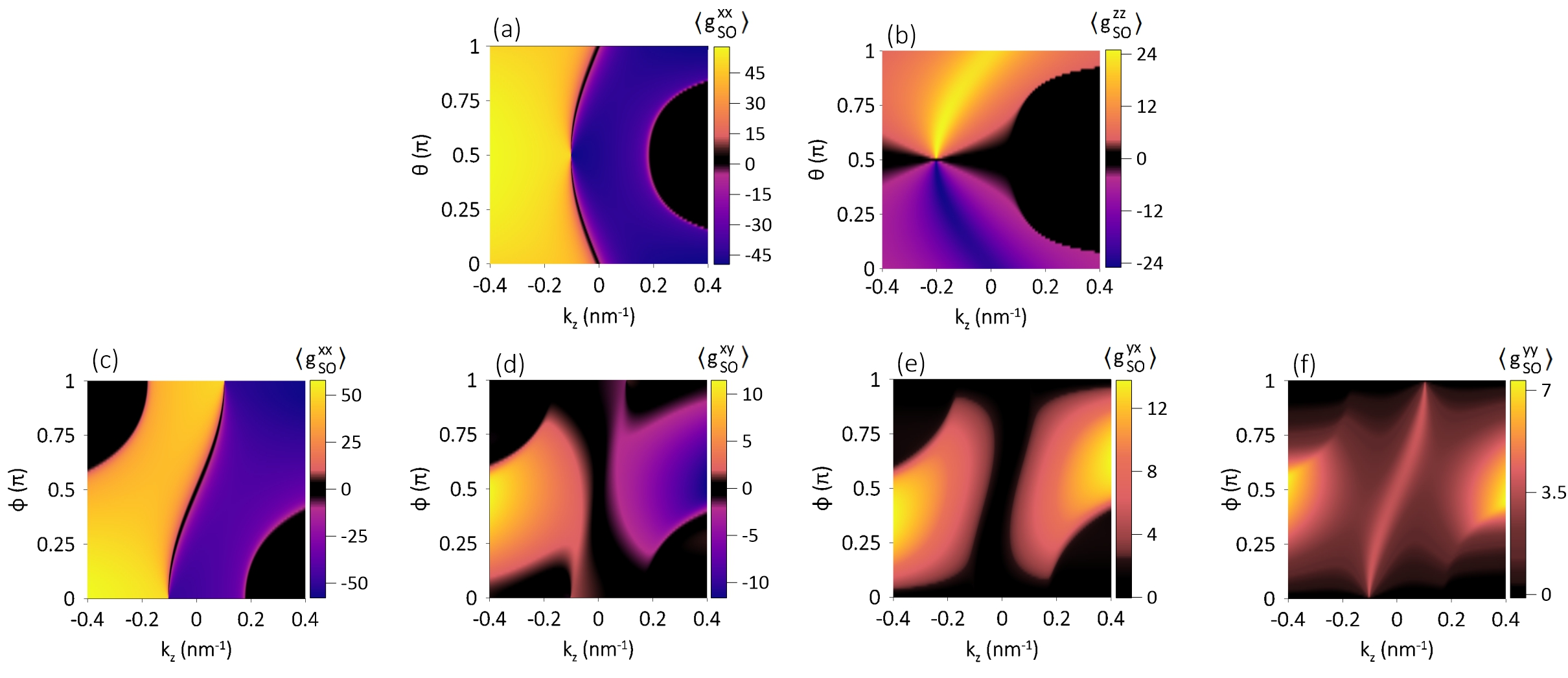}
    \caption{Maps of tensor elements $\langle g^{ab}_{SO} \rangle$ ($a,b=\{x,y,z\}$) as a function of wave vector $k_z$ and magnetic field orientation when it is rotated
    (a), (b) in the $xz$ plane and  (c), (d), (e), (f) in the $xy$ plane. Results for $B=1$~T and $V_g=0.2$~V.}      
    \label{fig:Lande_anisotropy}
\end{figure*}

Figure \ref{fig:Lande_anisotropy_gstar} shows maps of $\gstar$ as a function of the wave vector $k_z$ and the rotation angle for three considered rotation plane of the magnetic field. The effective \Lande\ factor $\gstar$ determined at $k_z=0$ - see right polar plots in Fig.~\ref{fig:Lande_anisotropy_gstar} - exhibits the two fold anisotropy when the magnetic field is rotated in the $xz$ and $xy$ plane with the maximal value twice larger than $g_{RLZ}$ for the magnetic field aligned along the $x$ axis. The rotation in $yz$ plane does not significantly change $\gstar$ exhibiting nearly isotropic behaviour. Similarly, as in the previous section, the observed anisotropy can be explained as a combination of two phonomena: (i) the orbital effects coming from the SO term and (ii) the polarization of the spin state being a resultant of the Rashba SO coupling and the magnetic field.

\begin{figure}[!htp]
    \centering
    \includegraphics[width = 0.35 \textwidth]{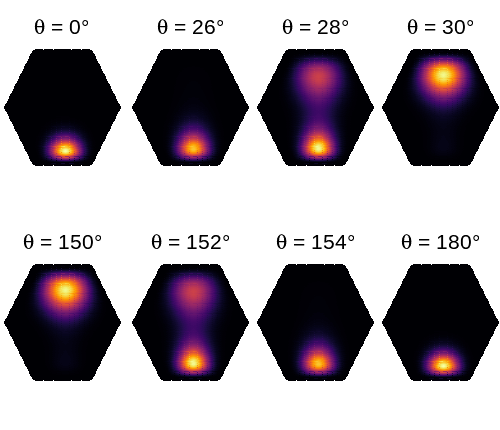}
    \caption{Squared envelope functions of the lowest subband for $k_z = 0.4$~nm$^{-1}$ as a function of $\theta$ at $\varphi = 0$ - rotation in the $xz$ plane. Note that the change of the wave function localization from the bottom to the top facet is quite abrupt and happens over an interval of $\sim 4^\circ$.}
    \label{fig:psi_theta_rotation}
\end{figure}

To get into details of the orbital contribution coming from SO coupling in Figs. \ref{fig:Lande_anisotropy}(a,b) we show maps of $\langle g_{SO}^{xx} \rangle$ and $\langle g_{SO}^{zz} \rangle$ as a function of the wave vector $k_z$ and $\theta$ when the magnetic field is rotated in the $xz$ plane. The black region on the right sides of both panels originates from the localization of the wave function far away from the bottom gate, in the region where the SO coupling is weak. This is apparent in Fig.~\ref{fig:psi_theta_rotation}, which shows the squared wave function for $k_z = 0.4$ nm$^{-1}$ under different magnetic field orientations.

\begin{figure}[!htp]
    \centering
    \includegraphics[width = 0.3 \textwidth]{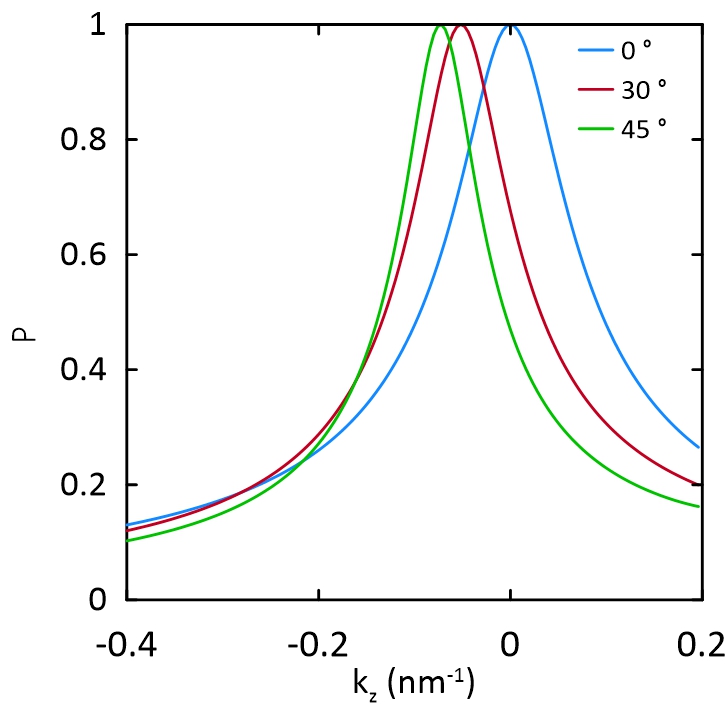}
    \caption{Spin polarization $P$ as a function of wave vector $k_z$  for the lowest subbands as a magnetic field $B= 1$~T is rotated in the $xz$ plane, at selected angles (see legend).}
\label{fig:Spin_ratios}
\end{figure}

Interestingly, we observe unusual behavior in the region where $\langle g_{SO}^{xx} \rangle$ changes sign. As discussed earlier, when the magnetic field is directed along the $x$-axis, this sign change is due to subband crossing. However, here the finite $z$-component of the magnetic field, perpendicular to the effective Rashba field, causes anticrossing of the subbands. The magnitude and position of the anticrossing in wave vector space depend on the orientation of $\mathbf{B}$. The behavior of $\langle g_{SO}^{xx}\rangle$ damping to zero at the sign change region, accompanied by a maximum in $|\langle g_{SO}^{zz} \rangle |$, can be explained by considering the evolution of electron spin at the anticrossing. Figure~\ref{fig:Spin_ratios} presents the $z$-spin polarization of the lowest subbands, defined as $P = \int  (|\psi^{\uparrow}_{k_z}(x,y)|^2 - |\psi^{\downarrow}_{k_z}(x,y)|^2) dxdy$, as a function of $k_z$ for different angles, $\theta$. We observe that at the anticrossing, the states become completely $z$-spin polarized, which maximizes $|\langle g_{SO}^{zz} \rangle|$. Simultaneously, the average value of $\sigma_x$, which determines $|\langle g_{SO}^{xx} \rangle|$ [see Eq.~(\ref{eq:el_gdiag})], becomes zero, which explains its vanishing for a specific $k_z$ vector.

The evolution of the SO-induced \Lande\ factor in the $xy$ rotation planes, the second for which we observe two fold anisotropy and depicted in Figs.~\ref{fig:Lande_anisotropy}(c-f), is in general a result of the interplay between the wavefunction localization, which is determined by orbital effects, and the electron spin direction, which is defined by both the SO interaction and the external magnetic field. It is worth noting that when the magnetic field has a component along the $y$ axis, the off-diagonal elements of the $\mathbf{g}_{SO}$ tensor may also contribute significantly to the effective \Lande\ factor - the magnitudes of $\langle g_{SO}^{xy(yx)} \rangle$ in Figs.~\ref{fig:Lande_anisotropy}(d,e) are comparable to those of the diagonal elements.
 
Although the maps of the $\mathbf{g}_{SO}$ tensor elements presented so far provide valuable information and offer a precise representation of the physical phenomena underlying the anisotropy of $\gstar$, it becomes challenging to directly compare them with results of recent experimental evidence. In experiments, the $k_z$ vector is often not well-defined, and what is typically obtained is an average value of $\gstar$ over all electronic states involved in the transport. For this reason we define the mean value of $\mathbf{g}_{SO}$ tensor elements averaged over all occupied states
\begin{equation}
    \overline {g}^{ab}_{SO} = \frac{\sum_{k_z} |\langle g^{ab}_{SO}(k_z) \rangle | f(E_{n=1,k_z} - \mu, T)}{{\sum_{k_z} f(E_{n=1,k_z} - \mu, T)}},
\label{mean_over_states}
\end{equation}
where $a,b=\{x,y,z\}$.
Such an approach has been recently used for analyzing the SO coupling in NWs and good agreement with experiments has been obtained.\cite{Escribano}

\begin{figure}[!htp]
    \centering
    \includegraphics[width = .35\textwidth]{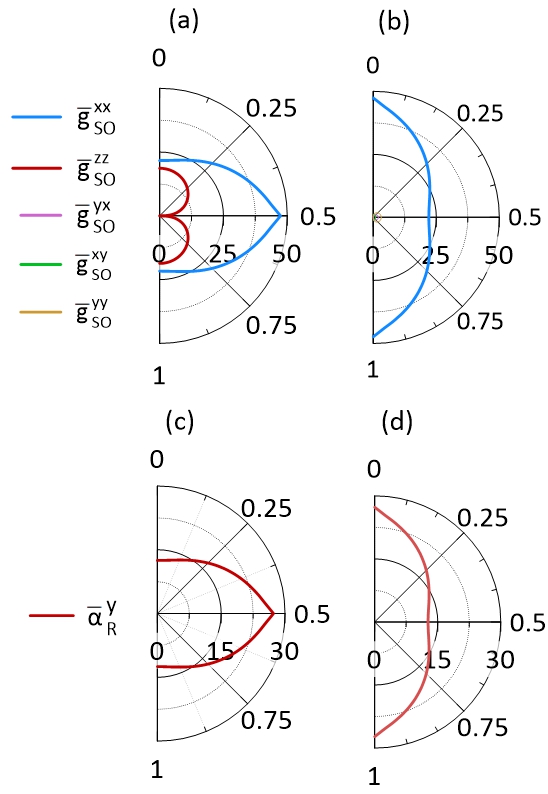}
    \caption{Averaged \Lande\ tensor elements $\overline{g}^{ab}_{SO}$ and the Rasha SO constant $\overline{\alpha}_R^y$ for the magnetic field rotated in three rotation planes (a,c) $xz$ and (b,d) $xy$. Results for the bottom gate potential $V_g = 0.2$~V and magnetic field $B = 1$~T.} 
    \label{fig:Lande_anisotropy_effective}
\end{figure}

In Fig.~\ref{fig:Lande_anisotropy_effective} we show the mean value of the tensor elements $\overline {g}^{ab}_{SO}$ and the Rashba SO constant $\overline{\alpha}^y_R$ (defined in the same manner) for the rotation planes $xz$ and $xy$ characterizing by the two-fold anisotropy of $\gstar$. We observe that irrespective of the rotation plane, all elements $\overline {g}^{ab}_{SO}$ exhibit strong anisotropy with a two-fold symmetry, closely corresponding to the evolution of the SO coupling, shown in Fig.~\ref{fig:Lande_anisotropy_effective}(c-d) (with a bottom gate $\overline{\alpha}^x_R=0$ due to the symmetry along the $y$ axis and it is not shown). A similar two-fold symmetry with respect to the magnetic field direction has been recently observed in the Rashba SO coupling measured for suspended InAs NW.\cite{Ioro} In both cases, the symmetry arises from the bottom gate architecture, which induces a large SO coupling near the bottom facet, while the rotating magnetic field alters localization of the wave function, due to the orbital effects. 

It is noteworthy that $\overline{g}_{SO}^{xx}$ remains the most robust against the rotation in the $xy$ plane [see Fig.~\ref{fig:Lande_anisotropy_effective}(b)], and it dominates over other terms for the considered gate setup. This can be attributed to the large coupling constant $\overline{\alpha}^y_R$ induced by the bottom gate voltage and the broken symmetry with respect to the $x$-axis -- see Eq.~(\ref{gxx}). Finally, it should be emphasized that the off-diagonal tensor components are one order of magnitude smaller than the diagonal ones. This observation holds true for the considered bottom gate configuration, which preserves symmetry around the $y$ axis, but it may differ for more sophisticated gate configurations as presented in the next subsection.

\subsection{Different gate configuration}

\begin{figure}[!t]
    \centering
    \includegraphics[width = .5\textwidth]{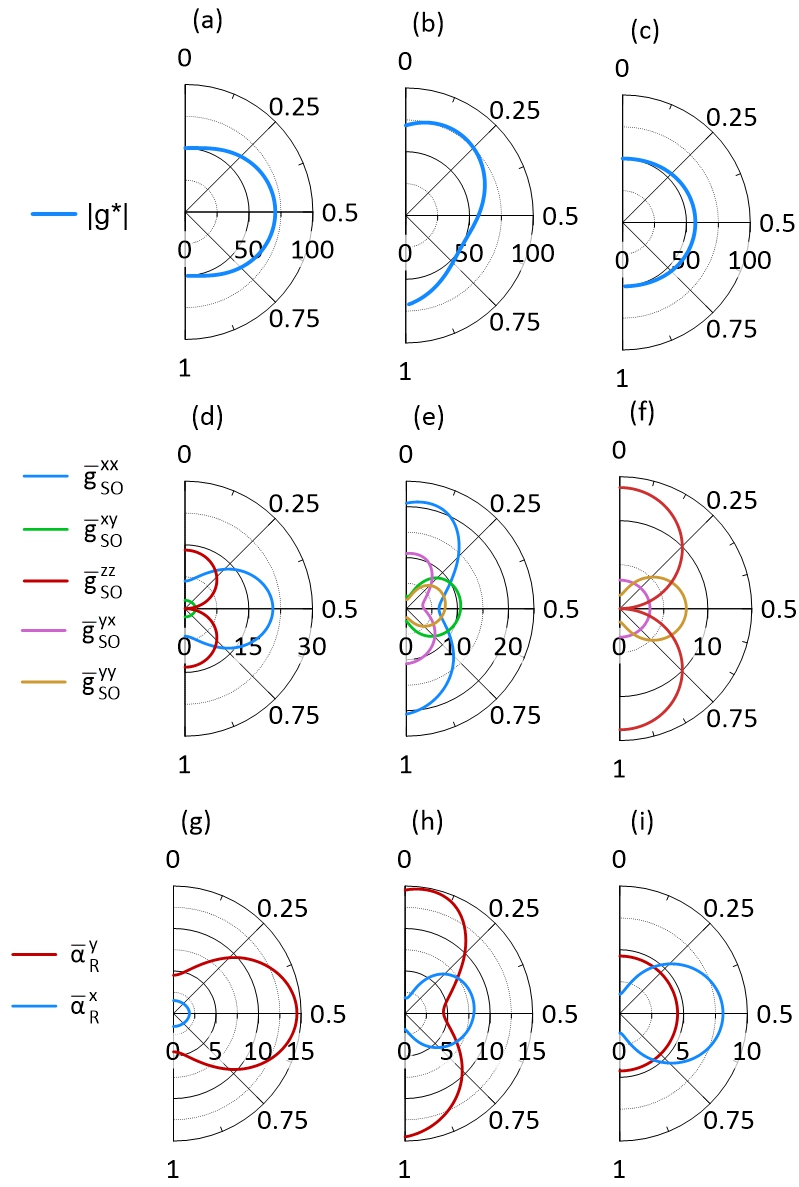}
    \caption{$\gstar$ [Eq.\ref{eq:g_energy}], \Lande\ tensor elements $\overline{g}^{ab}_{SO}$ and the Rasha SO constant $\overline{\alpha}_R^{x(y)}$ for the magnetic field rotated in three rotation planes (a,d,g) $xz$ (b,e,h) $xy$ (c,f,i) $yz$. Results for the gate configuration with the attached top and left-top gate and $V_g = 0.2$~V and magnetic field $B = 1$~T.} 
    \label{fig:Majorana_polar}
\end{figure}

In order to analyze in detail the magnitude of the off-diagonal elements of the $\mathbf{g}_{SO}$ tensor let us now consider an asymmetric gate configuration with two gates attached to the top and left-top facet. In this case the voltage applied to the gates generate both the $x$ and $y$ component of the Rashba SO coupling - see Fig.~\ref{fig:Majorana_polar}(g-i). In particular, the negative voltage generates the effective band bending near the gates similar to that observed in the Majorana NWs at the superconductor/semiconductor interface.\cite{Marcus2018} 

The $\gstar$ factor at $k_z=0$ is presented in Fig.~\ref{fig:Majorana_polar}(a-c). We see an enhancement of $\gstar$ with respect to $g_{RLZ}$ when the magnetic field is rotated in the $xz$ and $xy$ plane, with strong anisotropy determined by the gate configuration. As shown in Fig.~\ref{fig:Majorana_polar}(d), in this configuration, the off-diagonal elements of $\mathbf{g}_{SO}$ are of the same order of magnitude as the diagonal elements. This additional contribution plays a role in enhancing the overall effective \Lande\ factor. While the general principle that the largest SO-induced \Lande\ factor occurs when the magnetic field is perpendicular to $\boldsymbol{\alpha} _R$ is observed also for the this gate configuration. Consequently, we believe that our model, when applied to higher gate voltages, can account for the observed twofold enhancement of the effective \Lande\ factor, as recently observed in Majorana NWs.\cite{Schroer,vanWeperen2013,Marcus2018}

\section{Summary}

Based on the $\mathbf{k}\cdot\mathbf{p}$ theory within the envelope function approximation, we have analyzed the effective \Lande\ factor induced by the SO coupling in homogeneous semiconductor NWs under different magnetic field and gate configurations. By considering the orbital effects in the kinetic and SO terms, we have obtained the $\mathbf{g}_{SO}$ tensor which is treated as an auxiliary quantity to analyze the magnetic field dependence of $\gstar$. In the paper, we have studied the \Lande\ factor as well as the matrix elements of $\tgstar _{SO}$ with respect to the magnetic field magnitude and orientation.

We show that individual elements of the effective \Lande\ tensor induced by SO interaction are proportional to the Rashba coupling constant, which arises from the electric field generated by the adjacent gates. Hence, we have found that $\gstar$ is determined by two factors: 1) position and symmetry of the electron’s wave function, which can be tuned by the orbital effects, 2) the spin polarization of the electronic state. Specifically, when we apply the magnetic field perpendicular to NW, the inversion symmetry of the envelope functions is broken and the wave function is squeezed to the NW surface by a $k_z$-dependent effective potential. This effect results in an enhancement of $\gstar$ in a situation when the envelope function is squeezed to the facet near the gate where the electric field and consequently the Rashba SO coupling is larger. The opposite magnetic field (or $k_z$) results in the squeezing of wave function to the opposite facet where electric field from the gate and the corresponding SO coupling is weak, which results in nearly zero $\mathbf{g}_{SO}$ and $\gstar=g_{RLZ}$. 
On the other hand, for $\mathbf{B}$ directed along the NW axis the orbital effects are strongly reduced by the confinement and $\gstar$ as well as $\mathbf{g}_{SO}$ depends on the $z$ component of spin polarization, which is a resultant of the magnetic and effective Rashba field. Our results explains the recently demonstrated enhancement of the effective \Lande\ factor observed in semiconductor NWs as well as its anisotropy.\cite{Schroer,vanWeperen2013,Marcus2018}

Note that although our simulations have been limited to the regime where only the lowest subband is occupied, from our previous papers we expect that the electron-electron interaction, here introduced at the mean-field level, could be essential in estimating \Lande\ factor, via charge localization. At the high concentration regime total energy is minimized by reducing repulsive Coulomb energy, moving electrons outwards, and charge localizes at the six quasi-1D channels at the edges. As we discussed in Ref.~\onlinecite{Wojcik_anizotropy}, this strong localization is almost insensitive to the gate potential and the magnetic field direction.

Finally, we would like to underline that our model does not include the hole bands coupling expressed in the $\mathbf{k}\cdot\mathbf{p}$ model by the L\"uttinger parameters.\cite{Vezzosi2022} Note however, that as recently shown in Ref.~\onlinecite{Escribano} the applied conduction band approximation underestimates the SO coupling constant for the considered zinc-blende crystal structure. As the considered SO induced \Lande\ factor depends on the Rashba SO constants, we expect that the renormalization of the effective $\gstar$ observed in the experiments should be even greater than predicted by our results.

\section{ACKNOWLEDGEMENT}
The work was supported in part by PL-Grid Infras-
tructure, grant no. PLG/2022/015712.

\appendix*
\section{Dispersion relations}
\begin{figure*}[!htp]
    \centering
    \includegraphics[width = .9 \textwidth]{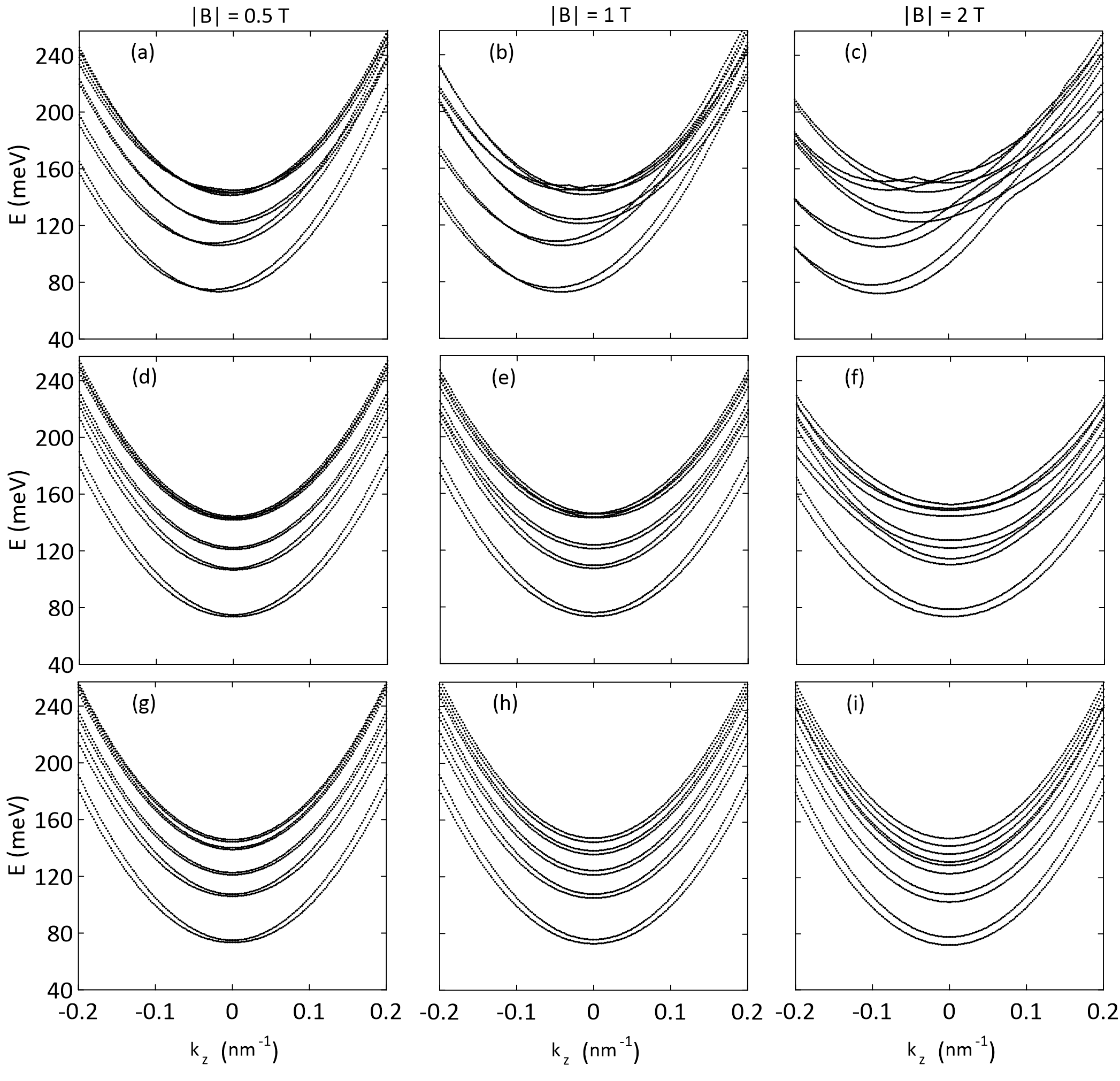}
    \caption{Dispersion relations $E(k_z)$ for ten lowest energy levels. Individual rows correspond to magnetic field directed along $x$-axis (a-c) , $y$-axis (d-f) and $z$-axis (g-i), respectively. Each column corresponds to the magnetic field magnitude: $|B| = 0.5$~T (a,d,g), $|B| = 1$~T (b,e,h) and $|B| = 2$~T (c,f,i).}      
    \label{fig:Dispersions_mozaique}
\end{figure*}
\begin{figure*}[!htp]
    \centering
    \includegraphics[width = .95 \textwidth]{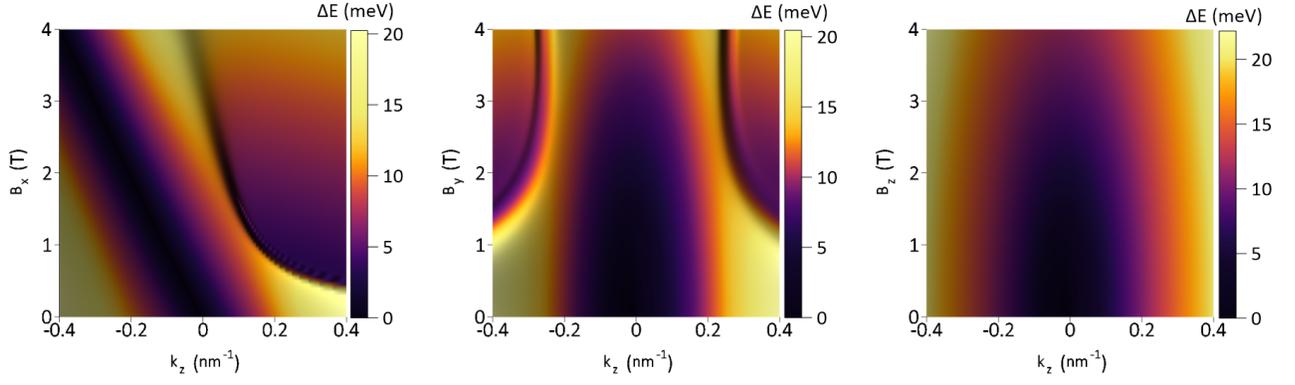}
    \caption{The energy difference between the first excited and ground state $\Delta E$ as a function of wavevector $k_z$ and the magnetic field applied along (a) $x$-axis, (b) $y$-axis, (c) $z$-axis. Results for $V_g = 0.2$~V.}      
    \label{fig:Splittings}
\end{figure*}

In the paper we have presented mainly $\gstar$, defined as the proportionality factor of the linear response of electronic states to the magnetic field. For completeness, the full dispersion relations $E(k_z)$ of the nanowire, including the interaction with magnetic field as well as the Rashba SO coupling, are presented in Fig.~\ref{fig:Dispersions_mozaique}, for chosen magnetic field magnitudes and directions. The corresponding maps presenting the energy difference between the first excited and ground state $\Delta E$ are presented in Fig.~\ref{fig:Splittings}.

\section{Size dependence}

\begin{figure}[!htp]
    \centering
    \includegraphics[width = .45 \textwidth]{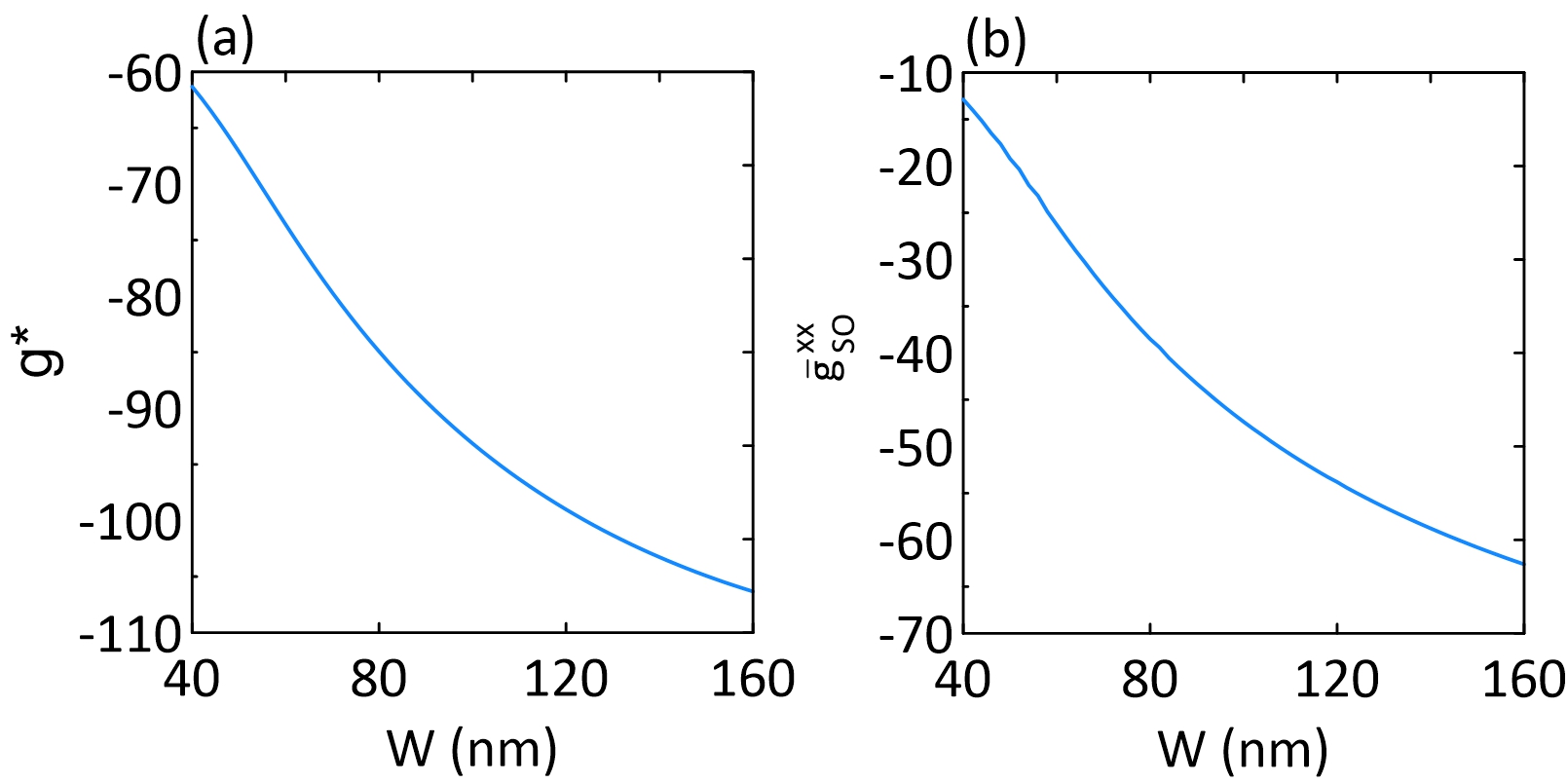}
    \caption{(a) $\gstar$ [Eq.~(\ref{eq:g_energy})] and (b) averaged $\overline{g}_{SO}^{xx}$ as a function of the NW width. Calculations for the magnetic field applied along $x$ direction, $B_x=0.1$~T and $V_g=0.2$~V. }
    \label{fig:size}
\end{figure}

Calculations presented in the paper have been carried out for the NW width $W=100$~nm for two reasons. First, it is a typical diameter of NWs fabricated by the commonly used fabrication methods and second, for this range of NW width, orbital effects considered here become significant. For completeness, in Fig.~\ref{fig:size} we present $\gstar (k_z=0)$ and $\overline{g}_{SO}^{xx}$ calculated with a magnetic field along the $x$ directions for which we observe the enhancement of the effective \Lande\ factor. As expected, for a small diameter, when the orbital effect are highly reduced, the SO induced \Lande\ factor approaches zero, which shows that the predicted enhancement of $\gstar$ is observable only for NWs of moderate or large width.

%\bibliography{refs.bib}
%merlin.mbs apsrev4-1.bst 2010-07-25 4.21a (PWD, AO, DPC) hacked
%Control: key (0)
%Control: author (8) initials jnrlst
%Control: editor formatted (1) identically to author
%Control: production of article title (-1) disabled
%Control: page (0) single
%Control: year (1) truncated
%Control: production of eprint (0) enabled
%

\end{document}